\input harvmac

\Title{LA-UR-96-00}{}

{\vbox{\centerline{THE GLUEBALL;}\vskip2pt\centerline
{THE FUNDAMENTAL PARTICLE OF NON-PERTURBATIVE QCD}}}


\bigskip\centerline{Geoffrey B. West\footnote{$^\ddagger$}{(gbw@pion.lanl.gov)}}
\centerline{High Energy Physics, T-8, MS B285}
\centerline{Los Alamos National Laboratory}\centerline{Los Alamos, NM  87545}\centerline{U. S. A.}



\vskip 1.5in

\centerline{\bf ABSTRACT}
\smallskip
{Theoretical ideas related to the existence of glueballs in QCD are reviewed. These include non-perturbative phenomena such as confinement, instantons, vacuum condensates and renormalons. We also discuss glueball dominance of the trace of the stress-tensor, the mass content of the nucleon and a theorem on the lightest glueball state.

\Date{8/96}

\vfill\eject

Glueballs are perhaps the most dramatic and novel prediction of QCD. From the vantage point of
twenty years ago when QCD was first being proposed as the fundamental theory of the strong
interactions, the idea that there might be quarkless hadronic states whose constituents were massless
gauge bosons (i.e. gluons) was almost revolutionary. Glueballs are inherently quantum
chromodynamic in nature and, as such, their existence is closely related to other essentially
non-perturbative phenomena that dominate low-energy hadronic physics such as the existence of vacuum
condensates and the dominance of glue in determining the gravitational mass of visible matter. They
clearly play a central role in elucidating QCD and their discovery would certainly be of great
significance. Indeed had such particles been found 15-20 years ago, their dicoverers would
certainly have been prime candidates for a Nobel Prize. Unfortunately, however, no
unambiguous experimental signal for their existence has thus far been found. This is due in
large part to the fact they can readily mix with ordinary quark model states and so can only be
identified by a process of elimination, i.e. by searching for extra states beyond conventional
``naive" quark model ones which have the correct decay characteristics. There has recently been a 
renewed flurry of interest, both experimental and theoretical, in these very interesting states and
the situation is, in fact, beginning to clarify 
\ref\rlone {For a review of recent experimental results and phenomenological interpretations,
see N.A. Tornquist "Summary of Gluonium95 and Hadron95 Conferences", University of Helsinki preprint
HU-SEFT-R-1995-16a, hep-ph/9510256.}%
\nref\rltwo {C. Amsler et al., Phys. Lett. {\bf B355}, 425 (1995).}%
\nref\rlthree {C. Amsler and F. Close, Phys. Lett. {\bf B353}, 385 (1995).}%
\nref\rlfour {T.Schaffer and E. V. Shuryak, Phys. Rev. Lett. {\bf 75}, 1707, (1995).}%
\nref\rlfive {J. Sexton, A. Vaccarino and D. Weingarten, Phys. Rev. Lett. {\bf 75}, 4563, (1995).}%
\nref\rlsix {V.V. Anisovitch and D.V. Bugg, ``Search for Glueballs", St. Petersburg preprint
SPB-TH-74-1994-2016.}%
--\ref\rlseven {A. Szczepaniak et al., Phys. Rev. Lett. {\bf 76}, 2011, (1996).}. 
Much detailed analysis has been performed on a large amount of experimental data with the
result that a few rather good candidates have emerged particularly in the region 1.5-1.7GeV
\rlone\rltwo\rlthree . In spite of this, however, the situation still remains unresolved and and more
work needs to be done. 
 
The theoretical situation is similarly somewhat ambiguous. Potential, bag and instanton gas models do
indeed indicate that the lowest state should be a scalar (and not a pseudoscalar or tensor, for
example) and that its mass should be in the above range \rlfour\rlsix\rlseven\ref\rleight{M.
Chanowitz and S. Sharpe, Nucl. Phys. {\bf B222}, 211, (1983).}. All of these models, in spite of
having the virtue of incorporating the correct low energy physics of  QCD, are only effective
representations of the full theory, and so their accuracy is difficult to evaluate. Recent intensive
lattice simulations of QCD focussed explicitly on the glueball are in general agreement with the
results of these models \rlfive . On the other hand, estimates from QCD sum rules indicate that the
pseudoscalar rather than the scalar should be the lowest state albeit with a mass also in the general
range of 1.5GeV \ref\rlnine {S. Narison, Z. Phys. {\bf C26}, 209, (1984) and private communication
\semi S. Narison and G. Veneziano, Int. J. Mod. Phys {\bf A4, no.11}, 2751, (1989).}. In addition 
there are field theoretic models in which the $2^{++}$ tensor is the lightest state\ref\rthree {M. Schaden and D. Zwanziger, "Glueball Masses from the Gribov Horizon", New York University preprint NYU-ThPhSZ94-1.}. This disagreement
between the QCD sum rules and the lattice estimates is surprising since these ought to be the least
model dependent and therefore the most reliable. However, the lattice simulations do use a quenched,
or valence, approximation, though it is generally believed that this is not a major source of error,
and the QCD  sum rules have difficulty satisfying a low energy theorem. Below I shall
prove a theorem that shows that, regardless of the model or approximation used, QCD requires that the
scalar must, in fact, be the lightest glueball state. As a corollary various mass inequalities such
as ${M(2^{++})\geq M(2^{-+})}$ can also be proven. 

Most of this paper will be devoted to a general overview of some of the theoretical ideas that
impact the glueball question and its relationship to QCD. I shall try to emphasise some issues and
developments that have not received quite as much attention in this context as some of the more
well-known topics such as quark and bag models, lattice gauge theory and so on. Among the topics that
I shall address are the operator description of the states, low energy theorems, glueball dominance
of the stress-energy tensor and its relationship to the gluon dominance of the proton mass. The
self-interaction of the gluons reflects the non-abelian gauge character of QCD; this is the origin
of both the possibility that there are glueball states as well as of the phenomenon of asymptotic
freedom. The latter is a property of the perturbative sector of the theory whereas the former is a
product of the non-perturbative. Furthermore, both of these remarkable phenomena arise in the
purely gauge sector of QCD and do not require the existence of quark degrees of freedom. Since
glueballs are inherently non-perturbative in nature their existence is closely related to color
confinement and to the existence of vacuum condensates and instantons. It is in this sense that they
can be dubbed the ``fundamental particles" of non-perturbative QCD. Ultimately one would like to be
able to start with the QCD Lagrangian and derive its spectrum in some well-defined approximation
scheme. Thus far this has proven impossible in spite of ambitious attempts such as the large $N_c$
expansion, chiral perturbation theory, soliton models, heavy quark expansions, instanton gas models
and so on. Apart from some recent work on the latter \rlfour\ these methods focus on the quark sector and have
had little to say about the glueball spectrum. Only lattice gauge theory \rlfive\ and, to some extent,
the sum rule consistency relations \rlnine\ can be said to have provided some direct contact with
fundamental QCD. Otherwise most of our intuition and predictions about glueballs are derived from
models.

Within the field theoretic framework of QCD all hadronic states are created by composite
operators constructed out of fundamental quark and gluon fields. Some well-known examples are the
following: 
$$\eqalign{\hbox{Scalars}\qquad {\sigma}_a(x) &\propto {\bar q}(x){\lambda}_a q(x)\cr
         \hbox{Pseudoscalars}\qquad {\phi}_a(x) &\propto {\bar q}(x){\gamma}_5 {\lambda}_a q(x)\cr
        \hbox{Vectors}\qquad {\rho}_a^{\mu}(x) &\propto {\bar q}(x){\gamma}^{\mu} {\lambda}_a q(x)\cr
          \hbox{Glueball}\qquad G(x) & \propto {F_{\mu\nu}^a(x)F_a^{\mu\nu}(x)}\cr
        \hbox{Glueball}\qquad \tilde G(x) &\propto {F_{\mu\nu}^a(x)\tilde F_a^{\mu\nu}(x)}}$$

By analogy with the ordering of operators in the operator product expansion it is natural to
order these by dimension. It was originally suggested by both Bjorken and Jaffe et al.
\ref\ten{J.D.Bjorken, Proc. Summer Institute on Particle Physics, SLAC, 1979; R.L.Jaffe, K.Johnson and
Z.Ryzak, Ann. Phys. {\bf 168}, 344, 1986} that, at least heuristically, one might expect the mass of
a state to increase with the dimension of the corresponding lowest dimensional operator that can
produce it. In the table below an obvious shorthand is used to describe the operators: $\Gamma$
represents a gamma matrix, $D$ the covariant derivative and $F$ the gluon field tensor. The most
salient feature of this is that all of the conventional quark model states are indeed those of lowest
dimension while the exotic states, namely the glueball, hybrid and ``molecular-like" ones are of
higher dimension. Though suggestive this does not explain why the quark model states should so
dominate the low energy spectrum. Notice also that there are many states with the same quantum
numbers arising from quite different operators leading to the complication of untangling the ``pure"
states from the physical states. On the other hand the lowest hybrid operator does give rise to a
state which cannot occur in the quark model, the $1^{-+}$. An unambiguous discovery of such a state
in the low energy spectrum would indeed have been a major triumph for QCD.

$$\halign{\qquad\qquad\hfil#&\qquad\hfil#\hfil&\qquad\hfil#\hfil&\qquad#\hfil\cr
Dimension&Operator&$J^{PC}$&Character\cr
3&{$\bar q\Gamma q$}&${0^{-+},1^{--},0^{++},1^{+-},1^{++}}$&QuarkModel\cr
4&{$\bar q\Gamma Dq$}&${2^{++},2^{-+},2^{--}}$&QuarkModel\cr
4&{$F^2$}&${0^{++},0^{-+},2^{++},2^{-+}}$&Glueball\cr
5&{$\bar q\Gamma Fq$}&${0^{-+},1^{-+},0^{++},2^{-+}}$&Hybrid\cr
6&{$F^3$}&${0^{++},0^{-+},1^{+-},3^{+-}}$&Glueball\cr
6&{$\bar q\Gamma q\bar q\Gamma q$}& 0 &``Molecules"\cr}$$

To understand somewhat more quantitatively why glueballs, for example, should have a higher mass
than a typical light quark state it is useful to use the language of a potential or bag model. The
argument I shall present is very simple and shouldn't be taken too seriously though it is useful for
giving some insight into what the important physics is at work here \ref\eleven{This argument was
developed following a stimulating discussion with Abe Seiden.}. There are many variants of the
color-force potential but all of them have two major characteristics in common corresponding roughly
to the perturbative and non-perturbative aspects of the theory: a Coulomb-like piece and a long-range
confining piece. A simple qualitative representation is  \eqn\one{V(r) = -{\alpha\over r} + \sigma r} 
where $\alpha \approx 0.2$ and $\sigma$ (the string tension) $\approx 400MeV$. In QCD there is, of
course, only a single scale parameter, namely the running coupling constant $\alpha_s(\mu)$ defined
at some scale $\mu$. All of the parameters of an effective potential, such as  $\alpha$ and $\sigma$
occurring in eq. \one , are, in principle, expressible in terms of $\alpha_s(\mu)$. In the simplest
version of the quark model this potential is used in a Schrodinger equation with
quarks whose effective mass is roughly 300MeV. One of the great mysteries of QCD is that this
prescription gives a remarkably good accounting of the low-lying hadrons. In QED (the limit 
$\sigma = 0$, $\alpha = e^2$ in eq. \one) the total energy is given by 
\eqn\two {E = {p^2\over 2m} - {e^2\over r}}
where $p$ is the momentum and $m$ the mass. From the uncertainty principle $pr\geq 1$, so
\eqn\three {E \geq {1\over 2mr^2} - {e^2\over r}}
Minimising this lower bound gives $E_{min} = -me^2/2$ with $r_{min} = 1/{me^2}$ which agree with the 
ground state values for the hydrogen atom. Let us apply this to the glueball considered as a bound
state of two massless gluons:
\eqn\four {E = 2p + {9\over 4}\sigma r - {\alpha\over r}}
The factor $9/4$ is simply a color factor. Minimising as before leads to 
$r = 2/3[(2 -\alpha)/\sigma]^{1/2}$ and  
$E = 3[(2 -\alpha)\sigma]^{1/2}\approx 3\sqrt 2\sigma^{1/2}\approx 1.7 $GeV. Not surprisingly this
shows that the glueball mass is governed by the non-perturbative string tension. Furthermore, even 
though $\sqrt\sigma\approx 400$MeV sets the scale, it also shows that the expected mass of the
lightest glueball is quite large, between 1.5 and 2GeV.  A similar calculation can be performed for a
typical meson. The analog to eq. \four\ is   
\eqn\five {E = 2(p^2 + m^2)^{1/2} - 2m +\sigma r - {\alpha\over r}} which leads to $E_{min}\approx
750$MeV. It is also possible to extend this argument to hybrids by considering them as bound states
of two massive quarks and a massless glueball; similar calculations to the above indicate that their
lowest mass is in the range of 2.5GeV. This argument therefore shows that glueballs should be heavier
than light quark states but lighter than hybrids.

The discretized version of these composite operators (or a smeared out version of them) is what is
used in lattice gauge theory to simulate the behaviour of the corresponding propagators thereby
allowing a ``measurement'' of the relevant hadronic mass. As already remarked there has been a
significant amount of work done using this approach to study the glueball. The most intensive study
\rlfive reveals that the $0^{++}$ should have a mass of approximately 1.7GeV somewhat higher than
those considered to be the best experimental candidates (at approximately
1.5GeV) \rlone\rltwo\rlthree ; however, these are within experimental (and presumably theoretical!)
limits.

 Before discussing QCD sum rules,
instantons and the like it is worth digressing here to emphasise the special role played by
the glueball in QCD beyond that of the ``hydrogen atom of non-perturbative physics''. Recall first
that the glueball field 
\eqn\six{G(x) = f_{G}F_{\mu\nu}^a(x)F_a^{\mu\nu}(x)}
is identical, up to constant factors, to the Lagrangian density of the pure gauge sector.
Furthermore, it is also identical to the trace of the stress-energy tensor, $\theta$,
which is the operator that determines masses of particles. The renormalisation of the trace
anomaly in the  triangle graph occurring in the $\theta gg$ vertex leads to
\eqn\seven {\theta = \sum m_q \bar q q + {\beta (g)\over g}F^2_{\mu\nu}}
where $\beta (g)$ is the conventional $\beta$ function: $\beta (g) = -bg^2 + \dots$ with 
$b = (11 - 2n_f)/48\pi^2$. Thus, even in massless QCD hadrons can be massive since $\theta \not= 0$.
Indeed, eq. \seven\ naturally leads to the idea of ``glueball dominance of the trace of the stress
tensor'' (at least when quark masses can be neglected): 
\eqn\eight{\theta (x) = f_G  m_G^2 F^2_{\mu\nu}(x) = m_G^2 G(x)}
Notice that $f_G m_G^2 = \beta (g)/ g$. Eq. \eight\ is the exact analog of both PCAC 
($\partial_\mu A_\mu = f_\pi m^2_\pi \phi_\pi$) and vector dominance of the electromagnetic current
($J_\mu = f_\rho m^2_\rho \rho_\mu$). By taking matrix elements of \eight\ between hadronic (H) states
at rest and using the fact that
\eqn\nine {\langle p|\theta|p\rangle = M (Baryons) ;\quad 2m^2 (Mesons)}
Goldberger-Treiman type relations can be derived \ref\one{V.A.Novikov et al., Nucl.Phys. B {\bf
191}, 301, (1981); M. Shiffman, Z. Phys. {\bf{C9}},347, (1980).}. Generically, these are of the form $f_{G} g_{GHH}\approx M_H$. The continuation in
mass to the physical region is quite severe here; however, this does allow a rough estimate of
hadronic couplings relevant to experimental searches. Since the stress tensor itself generates the
full Poincar$\acute e$ algebra and, in particular,  $\theta = \partial_\mu D_\mu$, where $D_\mu =
x^\nu \theta_{\mu\nu}$ is the dilation current which  is the generator of scale transformations, the
glueball is part of a rich algebra (akin to chirality) from which low energy theorems can be derived.
For example, one such theorem is  $f_G^2 m_G^2\approx 16\pi b\alpha_s E^4$ where
\eqn\ten{E \equiv \langle 0|G(0)|0\rangle}
is the energy density of the glueball vacuum condensate. 

Another interesting example is provided by the mass of the nucleon: since the masses of the u and
d quarks are only a few MeV and heavy quarks are not a major component of the nucleon almost all
of its mass must be derivable from the gluon field. Put slightly differently: if there were no
gluon component in eq. \seven\ the nucleon would weigh only a few MeV! 
Thus $M_N \approx (\beta (g))/ g\langle p|  F^2_{\mu\nu}|p\rangle$. This, in fact, is not quite right 
because heavy quarks can, in fact, contribute to \seven\ through a triangle graph which then connects
to the nucleon through gluons; (this is effectively the sea contribution)\ref\rthirteen{M. A. Shifman, A. I. Vainshtein and V. I. Zakharov, Phys. Lett {\bf{78B}}, 443, (1978); S. Raby and G. B. West, Phys. Rev. {\bf{D38}}, 3488, (1988)}. In the limit
$m_q\rightarrow\infty$ this gives 
\eqn\eleven{\langle p|\sum m_q \bar q q |p\rangle \approx 
                            -{n_h g^2 \over 24\pi^2}\langle p|F^2_{\mu\nu}|p\rangle}
where $n_h$ is the number of heavy quark flavours. This contribution exactly cancels the heavy quark
contribution in the $\beta$ function so $M_N \approx (\beta_l (g)/ g)\langle p| F^2_{\mu\nu}|p\rangle$
where the subscript $l$ indicates that only light flavours are to be counted in $\beta$. This is an
elegant example of the decoupling theorem at work. Because of eqs. \six\ and \eight\ this formula
explicitly exhibits glueball dominance in determining masses of light hadrons.

The role of the s-quark is ambiguous in this analysis since its mass is comparable to the perturbative
scale. Its contribution, $\langle p| m_s \bar s s |p\rangle$, can be estimated from the sum rule for
the nucleon sigma term and the Gell-Mann-Okubo formula for symmetry breaking of baryon masses. The
upshot of a careful analysis is that it contributes about $30\%$ of the mass, most of the rest being
from glue and only a few per cent actually being derived from the light u and d quarks\ref\rfourteen{X. Ji, Phys. Rev. {\bf{D52}}, 271, (1995)}}. This
situation is reminiscent of the ambiguities in interpretation of the origin of the nucleon spin and,
indeed, both the s-quark and a triangle anomaly play important roles in both analyses. The
``paradoxical'' nature of these problems can be highlighted by observing that, if one neglects the
s-quark contribution, then the nucleon mass can be expressed as $M_N = [(33 - 2n_l)/2n_h]\langle
p|\sum m_h \bar h h |p\rangle$ which would  seemingly imply that it is derived solely from its heavy
quark content! Of course the decoupling theorem obtained through the triangle graph shows that this
is, in fact, identical to the purely (low-energy) gluon contribution as in eq. \eleven . Care must
therefore be taken in how these formulae are interpreted. 

The mass of the glueball is determined by the leading singularity in its propagator which, if the
glueball is stable, is just a simple pole. Both the mass and the propagator satisfy renormalisation
group (RG) equations. Consider massless QCD, then the only scale in
the problem is the renormalisation scale, $\mu$, needed to specify the physical coupling, $g(\mu)$,
so, on dimensional grounds
\eqn\twelve{ m_G = \mu f[g(\mu)]}
Since $\mu$ is arbitrary, $dm_G/d\mu = 0$ which leads to the most elementary RG
equation
\eqn\thirteen {{d\ln f\over dg} = {1\over \beta (g)}}
and, therefore,
\eqn\fourteen {m_G = c_G\mu exp\int{dg\over \beta (g)}\equiv c_G\Lambda_{QCD} 
                                                         \approx c_G\mu e^{1/bg^2}}
where $c_G$ is a constant that determines the glueball mass in terms of $\Lambda_{QCD}$ . In the
second part of this equation the perturbative expansion for $\beta(g)$ has been used. Eq. \fourteen\
shows explicitly how mass can be generated in a massless theory (``dimensional transmutation'') and,
more significantly here, that it is is inherently non-perturbative. Notice, however, that this
non-perturbative behaviour is generated from perturbative effects via renormalisation and
characteristically leads to $e^{1/bg^2}$. This behaviour is called the renormalon contribution by
analogy with that of the instanton which has a characteristic $e^{8\pi^2/g^2}$ behaviour. Instantons
arise from non-trivial local minima of the action. For example, consider the scalar correlator    
\eqn\fifteen{\Gamma({\bf x},t) \equiv \langle 0|T[G({\bf x},t) G(0)]|0\rangle}
 which has a standard path integral representation \ref\rlGF' {Explicit gauge fixing terms have been 
suppressed since these do not affect the positivity of the measure in Euclidean space used below to
derive inequalities \rleleven ; if desired, a convenient and natural gauge choice is the axial one. In
addition, the sum over quark flavors is to be understood.}:  
\eqn\thirtythree{\Gamma({\bf x},t) =\int{{\cal D} A_{\mu}^a
e^{{i\over4}{\int F_{\mu\nu}^aF_a^{\mu\nu}d^4x}}}{det(\not{D} +m)}{G({\bf x},t) G(0)}}    
An expansion of its Fourier transform, $\Pi({q^2/ \mu^2},g^2)$, in terms of $g^2$ is generically of
the form:  \eqn\sixteen{ \Pi\bigg({q^2\over \mu^2},g^2\biggr) = \sum^\infty_{n=0}a_n(q^2)g^{2n}
                                     + \sum^\infty_{m,n=0}e^{-8\pi^2(m+1)/g^2} b_{mn}(q^2)g^{2n}} 
The first term represents ordinary perturbation theory (i.e. an expansion around the trivial vacuum
where the action vanishes) and the second an expansion around instantons whose action is an integral
multiple of $8\pi^2$.

A Kallen-Lehmann representation for $\Gamma({\bf x},t)$ can be inferred from asymptotic freedom and
the fact that $G(x)$ is of dimension 4:  
\eqn\seventeen{\Gamma({\bf x},t) = \Pi ^\prime (0,g^2) \partial^2 \delta^{(4)}(x) +\Pi(0,g^2)
\delta^{(4)}(x) + {\partial^4} \int\limits_{M_G^2}^{\infty}{dq^2\over{q^4}}\rho ({q^2/g^2,g^2})
{\Delta}_F (x,q^2)} 
Here $\rho (q^2)$ is the spectral weight function and $\Delta_F (x,q^2)$ the
standard free Feynman propagator. Correspondingly,
\eqn\eighteen{ \Pi\biggl({q^2\over \mu^2},g^2\biggr) = \Pi(0,g^2) + q^2 \Pi'(0,g^2)
      + q^4 \int\limits_{M_G^2}^{\infty}{dq'^2 \rho (q'^2/\mu^2,g^2)\over q'^4 (q'^2 - q^2)} } 
This dispersion relation and its implied high energy perturbative contribution is the starting point
for the QCD sum rule consistency conditions. The right-hand-side is saturated with known, or
presumed, resonances (the various glueball and quark mesonic states) and its high energy tail by a
perturbative contribution derived from asymptotic freedom. On the left-hand -side the operator
product expansion is used to express $T[G({\bf x},t) G(0)]$ in terms of a complete set of operators
of increasing dimension. In pure QCD this gives rise to a series with the (symbolic) structure:
\eqn\nineteen{ \Pi\biggl({q^2\over \mu^2},g^2\biggr) = b_1  \langle 0|F_{\mu\nu}^2|0\rangle
            + b_2  \langle 0|F_{\mu\nu}^3|0\rangle + b_3  \langle 0|F_{\mu\nu}^4|0\rangle + \dots}
where the coefficients $b_n$ are calculable. Masses of hadronic states are then related to the
vacuum condensates occuring in this equation; (the first of these is essentially $E$ of eq. \ten ).
For the glueball channel a detailed analysis has been carried out by Narison and Veneziano \rlnine\ who
concluded that the ground state is the $0^{+-}$ rather than the $0^{++}$ expected from naive
potential and bag models as well as from an intense lattice gauge simulation. As already remarked we
shall prove below that, at least in pure QCD, the $0^{++}$ must be the lightest state. Before doing
so it is worth remarking that the general constraints imposed on the propagator (and, therefore,
implicitly the mass) by the RG, analyticity and the existence of a perturbative regime are
non-trivial to satisfy \ref\rfifteen{G. B. West, Nuc. Phys. (Proc. Suppl.) {\bf{1A}}, 57, (1987)}. Roughly speaking, the RG forces $ \Pi({q^2/ \mu^2},g^2)$  to be a function of
the single variable $(q^2/\mu^2) exp\int{dg\over \beta (g)}$, rather than of the two variables $q^2$
and $g^2$ separately, as in a perturbative Feynman graph expansion. Thus, if it is analytic
in $q^2$ and there is a mass gap, it cannot be analytic in $g^2$ so the perturbative
expansion must diverge and be, at best, asymptotic. This suggests that there must be some subtle
interplay between the perturbative and non- perturbative, somehow ``mediated" by the renormalon
contribution. One might, therefore, be able to improve the sum rule
predictions by enforcing the RG constraint; effectively, this amount to including renormalon
contributions.

Let us now show that the lightest glueball must be the $0^{++}$. Consider the quantity (for $t>0$)
$$\eqalignno{Q(t) &{} \equiv \int{d^3x \Gamma({\bf x},t)}  &\eqnn\myeqne\myeqne\cr &{}
          = \sum_{N}{{| \langle 0|G(0)|N \rangle |}^2  \delta ^{(3)}({\bf{p_N}})e^{iM_{N}t}}  
                &\eqnn\myeqnf\myeqnf\cr} $$
where $M_N$ is the invariant mass of the state $|N\rangle$. The Euclidean version of this (effectively
given by taking $t\to i\tau$) implies that, when $\tau\to \infty$, 
\eqn\thirtyseven{Q_E(\tau) \equiv Q(i\tau) \approx e^{-M_0 \tau}}
where $M_0$ is the mass of the lightest contributing state. An analogous result can be derived from
the Euclidean version of eq. \seventeen\enspace for the asymptotic behaviour of the full correlator
when either $\tau$ or $|x|$ become large. Up to powers, this simply reflects the exponential decay of
${\Delta}_F (x,\mu^2)$ in the deep Euclidean region. This asymptotic behaviour in Euclidean space
forms the basis for extracting particle masses from lattice QCD  simulations \rlfive\ and will
similarly play a central role in our proof. There are a couple of points worth remarking about it
before proceeding. First, in pure QCD, where the scalar and pseudoscalar glueballs are expected to be
the lightest states in their respective channels, $M_0 = M_G$ or $M_{\tilde G}$. In the full theory,
however, the corresponding lightest states are those of 2 pions and 3 pions, respectively, and even
the lightest glueballs become unstable resonances. In that case $M_0 = M_{2\pi}$ or $M_{3\pi}$. On
the other hand, in the limit when $\tau$ becomes large, but remains smaller than
$\sim{2M_G/{\Gamma^2_G}}$, where $\Gamma_G$ is the width of the resonance, one can show that the
exponential decay law, eq. \thirtyseven , still remains valid but with a mass $M_0$ given by $M_G$
rather than $M_{2\pi}$; (a similar result obviously also holds for the pseudoscalar case). The point
is that, if there are well-defined resonant states present in a particular channel, then they can be
sampled by sweeping through an appropiate range of asymptotic $\tau$ values where they dominate,
since $\tau$ is conjugate to $M_N$ \ref\rlten {This and the closely related problem of mixing between
quark and gluon operators will be dealt with in a forthcoming paper. For the purposes of this paper,
glueballs are defined as those states created out of the vacuum by purely (singlet) gluonic
operators; see also C. Michael, Nucl. Phys. {\bf B327}, 515 (1989).}.

The basic inequality that we shall employ is that, in the Euclidean region, 
\eqn\thirtyeight{{(F_{\mu\nu}^a \pm  \tilde F_a^{\mu\nu})^2  \geq 0 } \qquad\Rightarrow\qquad
{f_G^{-1}G_E({\bf x},\tau) \geq \pm f_{\tilde G}^{-1}\tilde G_E({\bf x},\tau)}}
where $G_E({\bf x},\tau) \equiv G_E({\bf x},it)$.
The integral version of this will be recognised as the original basis for proving the existence of
instantons, to which we shall return below. Although this inequality holds for classical fields, it
can be exploited in the quantized theory by using the path integral representation, eq. \thirtythree
, in Euclidean space where the measure is positive definite. The positivity of the measure has been
skillfully used by Weingarten \ref\rleleven {D. Weingarten, Phys. Rev. Lett. {\bf 51}, 1830 (1983)
\semi E. Witten, {\it ibid} 2351(1983).} to prove that in the quark sector the pion must be the
lightest state. Here, when combined with the inequality \thirtyeight , it immediately leads to the
inequalities (valid for $\tau >0$)   
\eqn\forty{{{{f_G^{-2}}{\Gamma_E({\bf x},\tau) }\geq {f^{-2}_{\tilde G}}{\tilde \Gamma_E({\bf
x},\tau)}}}    \qquad\hbox{and}\qquad {{f_G^{-2}}{Q_E(\tau) } \geq {f^{-2}_{\tilde G}}{\tilde
Q_E(\tau) }}}   
By taking $\tau$ large (but $<{2M_G/{\Gamma^2_G}}$) and using \thirtyseven, the inequality
\eqn\fortytwo{M_G \leq M_{\tilde G}} 
easily follows. In pure QCD where these glueballs are isolated singularities, their widths vanish and
the limit $\tau\to \infty$ can be taken without constraint.

Although this is the result we want, its proof ignored the existence of the vacuum condensate 
$E$, eq. \ten . Since $E\not=0$ the vacuum is the lightest state
contributing to the unitarity sum so $M_0 = 0$ and the large $\tau$ behaviour of
$\Gamma_E({\bf x},\tau)$ is a constant, $E^2$, rather than an exponential. Thus, the inequalities
\forty \enspace are trivially satisfied for asymptotic values of $\tau$ since there is no condensate
in the pseudoscalar channel. It is, incidentally, the occurrence of $M_G$ in a sub-leading
asymptotic role masked by this constant condensate term that makes its extraction from lattice data
so challenging. To circumvent this problem it is clearly prudent to consider either the
derivative of $Q(t)$ or, more generally, the time or space evolution of $\Gamma({\bf x},t)$ since
these remove the offending condensate contribution. Although many of the
subtleties can be finessed by considering  ${\nabla}^2 \Gamma_E({\bf x},\tau)$ it is instructive to
first consider (for $\tau>0$) 
\eqn\fortythree{{\dot Q_E(\tau)}  
      = -\sum_{N}{{| \langle 0|G(0)|N \rangle |}^2  \delta ^{(3)}({\bf{p_N}})M_{N}e^{-M_{N}\tau}}}
The vacuum state clearly does not contribute to this so its large $\tau$ behaviour is, up to a
factor $-M_0$, just that of eq. \thirtyseven\ except that $M_0$ is now the mass of the lightest
contributing particle state. Now, (for $\tau >0$),   
\eqn\fortythreea{\Gamma_E({\bf x},\tau) 
                         = \langle 0|e^{H\tau}G_E(0)e^{-H\tau}G_E(0)|0\rangle}
which implies
\eqn\fortythreeb{{\dot \Gamma_E({\bf x},\tau)}
                        = -\langle 0|G_E({\bf x},\tau)HG_E(0)|0\rangle} 
where, in the last step, the condition  $H|0\rangle = 0$ has been imposed. Notice that,
whereas both $Q_E(\tau)$ and $\Gamma_E({\bf x},\tau)$ are positive definite, their time derivatives
are negative definite. Now, at the classical level $H$ is positive definite. We can therefore
repeat our previous argument by working in Euclidean space and combining the inequalities
\thirtyeight\enspace with a path integral representation for \fortythreeb\enspace to formally
obtain (for $\tau >0$) the inequalities
   
\eqn\fortyfive{ {{f_G^{-2}}{\dot \Gamma_E({\bf x},\tau)}} \leq
                                {{f^{-2}_{\tilde G}}{\dot {\tilde\Gamma}_E({\bf x},\tau)}}   
\qquad\hbox{and}\qquad {f_G^{-2}{\dot Q_E(\tau)} \leq {f^{-2}_{\tilde G}{\dot{\tilde
                                                                               Q}_E}(\tau)}}} 
The large $\tau$ limit then leads to   \eqn\fortyseven{{{ f_G^{-2}}M_G e^{-M_G\tau} \geq
                               {f^{-2}_{\tilde G}}}M_{\tilde G}e^{-M_{\tilde G}\tau}} 
from which \fortytwo \enspace follows even in the presence of condensates.
 
There are some subtle points in this argument that require clarification, in particular the
nature of the path integral representation for \fortythreeb\ and the question of the vacuum
energy contribution. These are best dealt with using the language and results of the transfer matrix
formalism used in lattice theory since this is directly formulated in the Euclidean region 
as a Lagrangian theory where the measure is positive definite. Rather than showing how this can be done here, we shall instead
circumvent these technical problems by considering the space rather than
time evolution of $\Gamma$. To this end consider  
\eqn\sevena{{\nabla}^2 \Gamma_E({\bf x},\tau) =
-\langle 0|G({\bf x},\tau) {\bf P}^2 G(0)|0\rangle} where ${\bf P} = {\bf E}_a \times {\bf B}_a$ is
the 3-momentum operator. The asymptotic behaviour of the full correlator can be deduced from from its
Kallen-Lehmann representation, eq. \eighteen . From this one finds that the large $\tau$ behaviour of
${\nabla}^2 \Gamma_E({\bf x},\tau)$ is, up to powers, again $e^{-M_G \tau}$. The path integral
representation of \sevena , in which ${\bf E}_a$ is replaced by ${\dot {\bf A}}_a$, is negative
definite so all of the previous arguments go through leading to the inequality \fortytwo . Notice
that, unlike the time derivative case, the vacuum energy presents no complication since ${\bf
P}|0\rangle \equiv 0$.

The extension of the above argument to the general case showing that the scalar must be lighter
than all other glueball states, can now be effected. Introduce an operator,
$T_{\mu\nu\alpha\beta\dots}(x)$, constructed out of a  sufficiently long string of
$F_{\mu\nu}^a(x)'s$ and $\tilde F_a^{\mu\nu}(x)'s$ that it can, in principle, create an arbitrary
physical glueball state of a given spin. Generally speaking a
given $T$ once constructed can, of course, create states of many different spins, depending on the
details of exactly how it is constructed. As a simple example consider the fourth-rank tensor
\ref\rlsixteen {For simplicity color indices as well the trace operator over color matrices ensuring
the singlet nature of the states have been suppressed.}  
\eqn\fiftythree{T_{\mu\nu\alpha\beta}(x) = F_{\mu\nu}(x)F_{\alpha\beta}(x)}     
which creates glueball states with quantum numbers $2^{++}$ and $0^{++}$. Now, in Euclidean space, the
magnitude of any component of $F_{\mu\nu}^a(x)$, or $\tilde F_a^{\mu\nu}(x)$, is bounded by the
magnitude of  ${{[F_{\mu\nu}^a(x)F_a^{\mu\nu}(x)]}^{1\over 2}}$. Hence, any single component of
$T_{\mu\nu\alpha\beta}(x)$ must, up to a constant, be bounded by $G(x)$:
\eqn\fiftyfour{T_{\mu\nu\alpha\beta}(x) \leq  f_{G}^{-1}G(x)}     This inequality is the analog of
\thirtyeight\enspace and so the same line of reasoning used to exploit that inequality when proving
\fortytwo\enspace can be used here. Following the same sequence of steps leads to the conclusion that
$M_G$ must be lighter than the lightest state interpolated by $T_{\mu\nu\alpha\beta}(x)$, from which
the inequality 
\eqn\fiftyfive {M(2^{++})\geq M(0^{++}) \equiv {M_G}}  
follows. It is worth pointing out that the pseudoscalar analog of this operator can be similarly
bounded thereby leading to the inequality  $M(2^{++})\leq M(2^{-+})$. This argument can be generalized
to an arbitrary $T_{\mu\nu\alpha\beta\dots} (x)$ since, again up to  some overall constants analogous
to $f_G$, it is bounded by some power $(p)$ of $G(x)$; i.e., for any of its
components, $T_{\mu\nu\alpha\beta\dots}(x)\leq {G(x)}^p$ . Now, the  operator ${G(x)}^p$ has the same
quantum numbers as $G(x)$ and so can also serve as an interpolating field for the creation of the
scalar glueball. The same arguments used to prove that this $0^{++}$ state is lighter than either the
$0^{+-}$ or the  $2^{++}$ can now be extended to the general case showing that it must be lighter
than {\it any} state created by {\it any} $T$; in other words, the scalar glueball must indeed be the
lightest glueball state.

Finally, we make some brief remarks about the conditions under which the bound is saturated. Clearly
the inequality \thirtyeight\ becomes an equality when 
\eqn\fiftysix {F_{\mu\nu}^a (x) = \tilde F_{\mu\nu}^a (x)} 
i.e. when $E_i^a(x)=B_i^a(x)$, which is also the condition that minimizes the action and signals the
dominance of pure instantons. In such a circumstance the scalar and pseudoscalar will be degenerate.
However, the proof of the mass inequality \fortytwo\ only required \thirtyeight\ to be valid at
asymptotic values of $|x|$. Thus, the saturation of this bound actually only rests on the weaker
condition that $F$ be self-dual in the asymptotic region where it must vanish like a pure gauge
field. Similarly, the saturation of the general inequality showing the scalar to be the lightest
state occurs when {\it all} components of $F_{\mu\nu}^a(x)$  have the same functional dependence at
asymptotic values of $|x|$. Although this is a stronger condition than required by the general
asymptotic self-dual condition \fiftysix , it is, in fact, satisfied by the explicit single instanton
solution that satisfies it. For instance, in $SU(2)$, 
\eqn\fiftyseven {F_{\mu\nu} (x)={4 {\lambda}^2\over{x^2 + {\lambda}^2}}{\sigma}_{\mu\nu}} 
Thus, the splitting of the levels is determined by how much the asymptotic behaviour of the
non-perturbative fields differ from those of pure instantons. This therefore suggests a picture in
which the overall scale of glueball masses is set by non-perturbative effects driven by instantons
(thereby producing the confining long-range force) but that the level splittings are governed by
perturbative phenomena.

\listrefs

\end